\documentclass[10pt, prd,aps,superscriptaddress,footinbib,twocolumn]{revtex4-2}
\usepackage{color}
\usepackage{amsmath}
\usepackage{amsthm}
\usepackage{amssymb}
\usepackage{stmaryrd}
\usepackage{graphicx}
\usepackage{mdframed}
\usepackage{paracol}
\usepackage{siunitx}
\usepackage{cancel}
\usepackage[unicode=true,
 bookmarks=true,bookmarksnumbered=false,bookmarksopen=false,
 breaklinks=false,pdfborder={0 0 1},backref=false,colorlinks=true]
 {hyperref}

\usepackage[default]{changes} 
\usepackage{xcolor} 
\setaddedmarkup{\textcolor{red}{#1}}         
\setdeletedmarkup{\textcolor{red}{\sout{#1}}} 

\hypersetup{
 linkcolor=magenta, urlcolor=blue, citecolor=blue, pdfstartview={FitH}, hyperfootnotes=false, unicode=true}

\makeatletter

\usepackage{amsfonts}\usepackage{tabularx}\usepackage{dcolumn}\usepackage{bm}\usepackage{graphicx}\usepackage{epstopdf}\usepackage{tikz}
\usepackage{times}
\usepackage{cleveref}
\hypersetup{urlcolor=blue}

\DeclareMathOperator{\Tr}{Tr}

\makeatother

\newmdtheoremenv{theo}{Theorem}

\begin{document}


\title{Optimal Form Factors for Experimental Proposals on Gravity-Induced Entanglement}

\author{Ziqian Tang}\thanks{These authors contributed equally}
\affiliation{Beijing Key Laboratory of Fault-Tolerant Quantum Computing, Beijing Academy of Quantum Information Sciences, Beijing 100193, China}
\affiliation{Bejing National Laboratory for Condensed Matter Physics, Institute of Physics, Chinese Academy of Sciences, Beijing 100190, China}
\affiliation{University of Chinese Academy of Sciences, Beijing 100049, China}

\author{Hanyu Xue}\thanks{These authors contributed equally}
\affiliation{Yuanpei College, Peking University, Beijing 100871, China}

\author{Zizhao Han}
\affiliation{Center for Quantum Information, IIIS, Tsinghua University, Beijing 100084, China}

\author{Zikuan Kan}
\affiliation{School of Physics, Renmin University of China, Beijing 100872, China}

\author{Zeji Li}
\affiliation{School of Integrated Circuits, Tsinghua University, Beijing 100084, China}

\author{Yulong Liu}
\email{liuyl@baqis.ac.cn}
\affiliation{Beijing Key Laboratory of Fault-Tolerant Quantum Computing, Beijing Academy of Quantum Information Sciences, Beijing 100193, China}
\affiliation{Bejing National Laboratory for Condensed Matter Physics, Institute of Physics, Chinese Academy of Sciences, Beijing 100190, China}
\affiliation{University of Chinese Academy of Sciences, Beijing 100049, China}

\date{\today}

\begin{abstract}
The interface between quantum mechanics and gravity remains an unresolved issue. Recent advances in precision measurement suggest that detecting gravity-induced entanglement in oscillator systems could provide key evidence for the quantum nature of gravity. However, thermal decoherence imposes strict constraints on system parameters. For entanglement to occur, mechanical frequency $\omega_m$, dissipation rate $\gamma_m$, environmental  temperature $T$, oscillator density $\rho$, and the form factor $\Lambda$—determined by the geometry and arrangement of oscillators—must satisfy a specific constraint. This constraint, intrinsic to the noise model, is considered universal and cannot be improved by quantum control. Given the difficulty in further optimizing $\omega_m$, $\gamma_m$, $\rho$, and $T$, optimizing $\Lambda$ can relax the constraints on these parameters. In this work, we prove that the form factor has a supremum of $2\pi$, revealing a fundamental limit of the oscillator system. We propose designs that approach this supremum, nearly an order of magnitude higher than typical spherical oscillators. This optimization could ease experimental constraints and bring quantum gravity validation based on gravity-induced entanglement closer to realization.
\end{abstract}

\maketitle


\section{Introduction}
The problem of interfacing quantum mechanics and gravity has long been an unresolved issue in physics. However, to this day, no empirical evidence of the quantum gravity effect has been observed. In recent years, advances in precision measurement technology have raised the possibility of detection of gravitational effects within massive quantum systems at lower energies in the foreseeable future \cite{Santos2017, Li2018, Westphal2021, Whittle2021, Liu2021, Youssefi2023, Fuchs2024, Bose2023, Carney2019, Gallerati2022}. Specifically, detecting gravity-induced entanglement (GIE) in these systems is expected to provide crucial empirical evidence for the quantum nature of the gravitational field \cite{Feynman1957, Bose2017, Bose2022, Marshman2020, Marletto2017, Miao2020, Plato2023, Kafri2014, Krisnanda2020, Qvarfort2020, Cosco2021, Weiss2021, Balushi2018, Matsumura2020, Biswas2023}. Among these approaches, GIE-based experiments in quantum mechanical oscillator systems entangled through central-potential interactions are among the primary candidate methods \cite{Kafri2014, Krisnanda2020, Miao2020, Qvarfort2020, Plato2023}.

Various experimental schemes have been proposed for such studies. However, a common challenge in these proposals is that thermal decoherence can destroy the GIE, imposing strict constraints on the system parameters necessary for the experiment. Across proposals like two free oscillators \cite{Kafri2014}, linearized optomechanics \cite{Miao2020, datta2021signatures}, levitated nano-systems \cite{Rijavec2021}, and modulated optomechanics \cite{Plato2023},  requirements align with the same inequality $2 \gamma_m k_B T \lesssim \hbar G \Lambda \rho$ when $k_B T \gg \hbar \omega_m$ where $\omega_m$ and $\gamma_m$ are the mechanical frequency and dissipation of the oscillator, $T$ is the system's environmental temperature, $\rho$ is density of the oscillator, and $\Lambda$ is a form factor related to the geometry and spatial arrangements of the two oscillators, which also appears in various gravitational measurement experiments, such as those involving resonant detectors \cite{Schmole2016, Westphal2021, Liu2021}. Inherently a property of the noise model, the inequality applies universally to experimental systems and cannot be improved with novel quantum control techniques \cite{Miao2020, Plato2023}. 

This inequality reveals several significant challenges. First, it involves only four system parameters $\gamma_m$, $T$, $\rho$, and $\Lambda$, which restricts the system's tuning to these parameters alone, leaving little flexibility to optimize other parameters in order to satisfy the inequality. Furthermore, optimization becomes prohibitively costly when these system parameters reach specific thresholds. For instance, in large-mass mechanical oscillators, ground-state cooling introduces a theoretical limit to the effective temperature. Meanwhile, improvements in parameters such as dissipation $\gamma_m$ are constrained by the intrinsic material properties of the oscillator, offering little scope for further optimization through external means, e.g., applying higher prestress in membrane-based systems \cite{liu2025degeneracy}. Moreover, the maximum possible density, $\rho$, is limited by the densest known material, osmium, which has a density of approximately $22.4\ \si{g/cm^3}$. This restricts the upper bound on $\rho$, making it another challenge to optimize the system. Given these difficulties, one might naturally consider whether improvements in the geometry and spatial arrangement of the oscillators could increase the value of the form factor $\Lambda$, thus indirectly reducing the need for optimization of the other parameters.

\begin{figure}[htp]
    \centering    
    \hspace*{-0.3cm} 
    \vspace*{0cm}
    \includegraphics[width=0.9\linewidth]{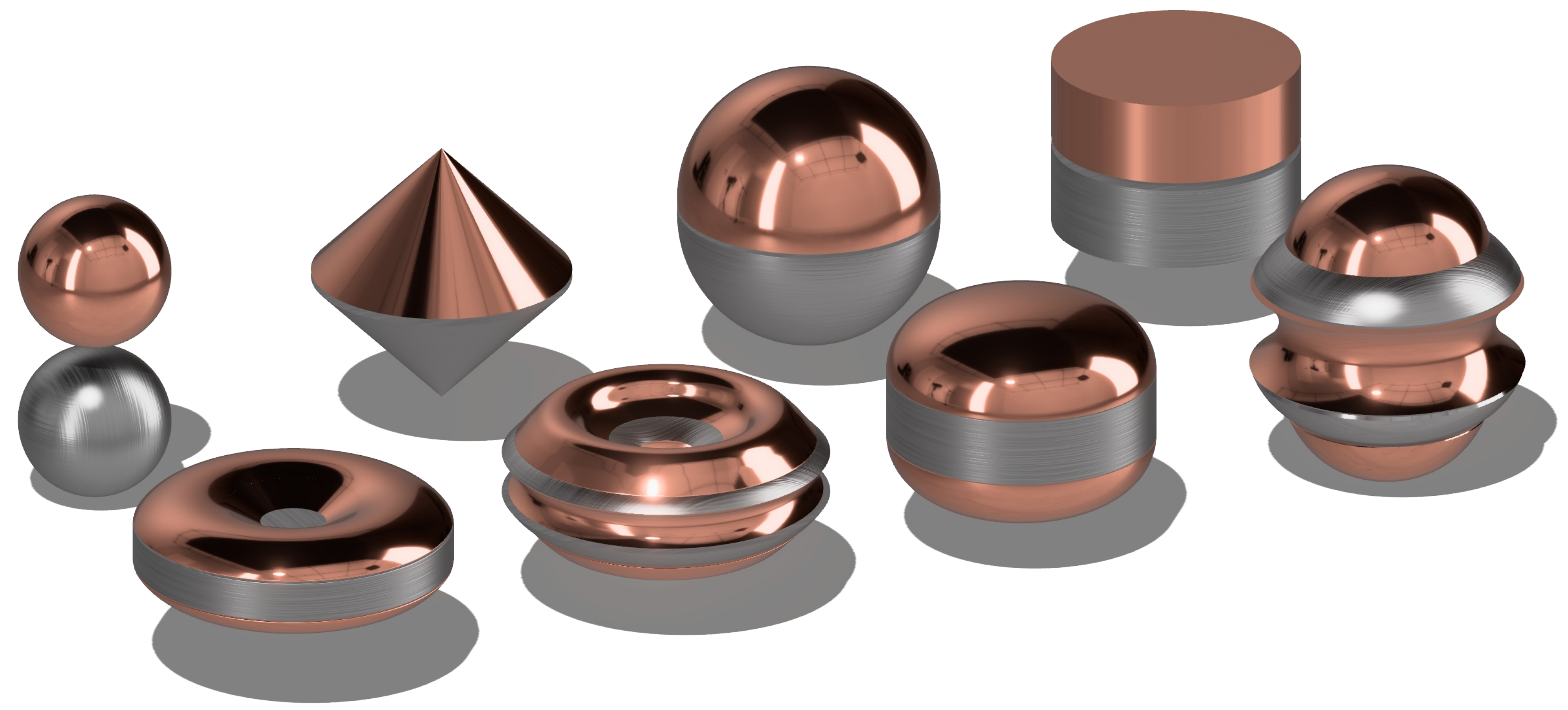}
    \caption{This figure presents several configurations with oscillation along the vertical axis. Colours distinguish the two masses and have no physical significance. The back row shows configurations of simple geometric shapes, while the front row includes fragmented configurations obtained via iterative optimization methods. In the front row configurations, each mass appears to be composed of several fragmented components, which can be understood as linked together through extremely thin bridges to form a connected whole. The $\Lambda$ values are $0.79$, $1.58$, $1.77$, $1.82$ (back row, left to right) and $1.95$, $2.67$, $2.81$, $3.33$ (front row, left to right). 
} 
    \label{fig:fig1}
\end{figure}

In a previous work, the case of two spherical oscillators was considered, for which it was shown that the form factor $\Lambda$ is bounded above by $\pi/3$ \cite{Krisnanda2020}. And for two oscillators arranged in a cylindrical configuration, the form factor can reach a value as high as $2.0$ \cite{Miao2020}, which is among the best in existing research. This naturally raises the question of whether other geometries and spatial arrangements could yield a higher value of $\Lambda$, and whether there exists a theoretical upper bound for $\Lambda$. Fig. \ref{fig:fig1} presents several typical configurations and their corresponding $\Lambda$ values.

In this work, a mathematical proof is provided that the supremum of $\Lambda$ over all possible geometries and spatial arrangements is $2\pi$, thus resolving this question substantially. The process of proving this result also inspired us in designing oscillators with larger form factors. The central idea is that the mass elements of the two oscillators should be intimately mixed rather than separated as two balls or cylinders. Moreover, the fine geometry of the mixture will strongly influence the form factor and should therefore be designed in a special way. We examined several specific oscillator configurations and verified through numerical calculations that, in certain cases, the form factors closely approach their theoretical supremum, nearing 6.0. This advancement may provide a substantive contribution to the experimental validation of quantum gravity effects.

\section{Optimal Form factor}
The form factor in oscillator-based GIE experiments derives from the gravitational interaction term in the Hamiltonian. For oscillators made of two identical natural frequencies $\omega_m$, masses $M$ with uniform density $\rho$ and arbitrary configuration – whether on a membrane \cite{Liu2021}, suspended by a pendulum \cite{Miao2020}, or as levitated nano-systems \cite{Rijavec2021} – the low-energy gravitational interaction Hamiltonian can be described by $\hat{H_G}\simeq \frac{\partial^2 V_G}{\partial x_1 \partial x_2}|_0 \hat{x}_1 \hat{x}_2$ \cite{Miao2020, Krisnanda2020, Plato2023}, where $V_G$ is the Newtonian potential between the two masses, $x_1$ and $x_2$ are displacements of them from equilibrium. Using the expression for the gravitational potential, the second-order partial derivatives can be represented as $\left|\frac{\partial^2 V_G}{\partial x_1 \partial x_2}|_0\right| = G M \Lambda \rho$, where $\Lambda$ is a dimensionless and scale-invariant quantity known as the form factor that depends solely on the geometry, spatial arrangements, and direction of oscillation of the oscillators \cite{Krisnanda2020, Miao2020}. 

To express $\Lambda$ explicitly, assuming that the two masses occupy separated domains $A$ and $B$, each with volume $V$, and are placed relative to each
other such that the relative displacement of the two masses is
always parallel to a fixed direction represented by a unit vector $\textbf{n}$. Then $\Lambda$ is given by
\begin{equation}
\label{eq:Lambda}
    \begin{aligned}
        \Lambda (A,B,\textbf{n}) = \left|\frac{1}{V}\int_A \int_B K_\textbf{n}(\textbf{r}-\textbf{r}') \mathrm{d}^3\textbf{r}\mathrm{d}^3\textbf{r}'\right|
    \end{aligned}
\end{equation}
where $K_\textbf{n}(\textbf{r}) = \textbf{n}^T K(\textbf{r})\textbf{n}$ is the kernel function with $K(\textbf{r})= -\nabla_\textbf{r}\otimes \nabla_\textbf{r}\frac{1}{r}$. Physically speaking, $\Lambda$ is actually the normalized tidal force in a specific direction $\textbf{n}$ between two masses.

In GIE experiments, for quantum entanglement to persist despite thermal decoherence, the interaction rate must exceed the decoherence rate, which yields $||\hat{H}_G|| / \hbar \gtrsim 2 M \gamma_m k_B T \delta x_q^2 / \hbar^2$ where $\delta x_q$ is a characteristic length scale given by \cite{Miao2020} and \cite{Braginsky1995}. With the expression of form factor $\Lambda$, this inequality can be re-expressed in the form $2\gamma_m k_B T \lesssim \hbar G \Lambda \rho$ which is a universal constraint on the parameters of GIE experiments. The inequality involves four experimental parameters: $\gamma_m$, $T$, $\rho$, and $\Lambda$. As mentioned, to satisfy the inequality, the importance of optimizing the form factor $\Lambda$ will become evident once the other parameters have been optimized to their limits. In existing studies, the form factor for specific geometries and spatial arrangements has been investigated. In \cite{Krisnanda2020}, the form factor for two spherical oscillators was studied, yielding \( \Lambda < \pi/3 \). In \cite{Miao2020}, the form factor for cylinders with different radii, heights and separation along the axis was studied, reaching a value of approximately 2.0. Next, we will present the limits of form factor optimization. 

\subsection{Supremum of form factor}
As the main result of this work, the supremum of the possible values of the form factor is provided for all geometries, spatial arrangements, and oscillation directions. To ensure the physical validity of the subsequent discussion, some additional constraints must be imposed on the domains $A$ and $B$ occupied by the two masses. First, $A$ and $B$ must be bounded, as it is physically impossible to create unbounded objects. In addition, $A$ and $B$ must be separated, so that each can be identified as distinct objects. Based on these considerations, the following theorem is obtained, which means that the optimization of the form factor has a supremum of $2\pi$:

\textit{Theorem 1.}–Let $\mathcal{S}$ be the set of pairs $(A, B)$ of three-dimensional bounded domains $A$ and $B$ of equal volume $V$, where $A$ and $B$ are separated. Denote by $S^2$ the unit sphere of all unit vectors $\textbf{n}$, then
\begin{equation}
    \begin{aligned}
        \sup_{(A,B)\in\mathcal{S}} \max_{\textbf{n}\in S^2} \Lambda (A,B,\textbf{n}) = 2\pi.
    \end{aligned}
\end{equation}

The complete proof of Theorem 1 will be included in the appendix. Instead, here we will further analyze the form factor $\Lambda$ from both the mathematical form and the physical meaning, providing a more interpretable path to understand the result of Theorem 1.

\subsection{Analysis of result}
The analysis of $\Lambda$ encounters some difficulties. Since the integral kernel $K_\textbf{n}$ is highly singular, $\Lambda$ is not even obvious to have an upper bound. To simplify the discussion while retaining generality, we assume the oscillation direction is along the $z$-axis, i.e., $\textbf{n}=\textbf{e}_z$. The integral kernel then becomes $K_{\textbf{e}_z}(\textbf{r})=\frac{x^2 +y^2 - 2z^2}{r^5}$. With this, a simple observation is that $\Lambda$ is scale-invariant: scaling both objects $A$ and $B$ by the same factor does not change $\Lambda$. Furthermore, $\Lambda$ increases as the distance between $A$ and $B$ decreases. Therefore, for two objects $A$ and $B$, a natural idea for designing their geometry to increase the form factor is to bring their centres of mass as close as possible. Based on this, a possible approach is to make $A$ and $B$ flat or elongated. This is also reasonable given the form of the kernel $K_{\textbf{e}_z}$, where the signs of the three terms in the numerator cancel each other out. To maximize $|K_{\textbf{e}_z}|$, one should reduce the cancellation by extending $A$ and $B$ along the $z$-axis or the $xy$-plane. Unfortunately, however, this approach is ineffective. For example, when $A$ and $B$ are infinitely large flat cylinders, one finds that $\Lambda$ decreases and tends to zero.

To further optimize $\Lambda$, an alternative approach is local geometric adjustments: introducing complementary grooves on the surfaces of $A$ and $B$, such as tooth meshing, can also bring them into closer contact, thereby increasing $\Lambda$.

In fact, one can imagine that the mass elements of the two objects $A,B$ are two piles of sand of different colours. They are \textit{mixed} together very tightly (with average density of every component $1/2$) but essentially \textit{separated}. On top of that, we can imagine a more thorough but different approach in which objects are \textit{superimposed} in space as the mixture appears to go from coarse to infinitely fine, leading to a uniform mass distribution with density $1/2$. In this case, $A,B$ should be considered as a single object $A\cup B$, occupying a volume of $2V$. The form factor of the superimposed state (denoted as $\Lambda_s$) then becomes
\begin{equation}
	\label{eq:LambdaSuperImposed}
	\begin{aligned}
	\Lambda_s (A\cup B,\textbf{e}_z)= \left|\frac{1}{4V}\int_{A\cup B} \int_{A\cup B} K_{\textbf{e}_z}(\textbf{r} - \textbf{r}')\mathrm{d}^3\textbf{r}\mathrm{d}^3\textbf{r}'\right|
	\end{aligned}
\end{equation}
Firstly, we focus on the superimposed state and prove $\sup\Lambda_s (A\cup B,\textbf{e}_z)=2\pi$. This is also the supremum of the case that $A,B$ are separated, which is of fundamental importance and is proved in the appendix. Because the integral kernel $K_{\textbf{e}_z}(\mathbf{r})=\frac{x^2 +y^2 -2z^2}{r^5}$ is very singular, the precise definition of the integral actually takes some effort. However, we get intuition from the similarity between Eq.~\ref{eq:LambdaSuperImposed} and the electric potential energy of an object uniformly polarized with polarization density $\textbf{P}$ along the $z$-axis. The latter is given by
\begin{equation}
	\begin{aligned}
		U= \frac{1}{8\pi\varepsilon_0}\int_{A\cup B} \int_{A\cup B} P^2 K_{\textbf{e}_z} (\textbf{r}-\textbf{r}')\mathrm{d}^3\textbf{r}\mathrm{d}^3\textbf{r}'
	\end{aligned}
\end{equation}
On the other hand, we have $U=\frac{1}{2}P\overline{E_z}$, where $\overline{E_z}$ is the average depolarization field inside the object. Also depends only on the geometry, the depolarization factor is defined by $\chi(A\cup B,\textbf{e}_z)={\varepsilon_0}\overline{E_z}/P$. Therefore, the form factor in the superimposed state is related to the depolarization factor by $\Lambda_s (A\cup B,\textbf{e}_z)=2\pi\chi(A\cup B,\textbf{e}_z)$. Because $\chi(A\cup B,\textbf{e}_z)$ gets its maximum $1$ when $A\cup B$ is an infinitely large plate extended in the $x$ and $y$ direction, $\sup\Lambda_s (A\cup B,\textbf{e}_z)=2\pi$. 

We can also prove it directly by considering $\Lambda_s (A\cup B,\textbf{e}_z)$ as the diagonal entry of a tensor $T(A\cup B)= \frac{1}{4V}\int_{A\cup B} \int_{A\cup B} K(\textbf{r}-\textbf{r}')\mathrm{d}^3\textbf{r}\mathrm{d}^3\textbf{r}'$. The tensor $T(A\cup B)$ is positive-definite because its eigenvalues correspond to the total energy of the electric field which is positive. Since $\Tr K(\textbf{r} - \textbf{r}') = 4\pi \delta(\textbf{r} - \textbf{r}')$, $\Tr T(A\cup B)$ is the integral of a delta function and equals $2\pi$. Only when $\chi(A\cup B,\textbf{e}_x)=\chi(A\cup B,\textbf{e}_y)=0$, $\chi(A\cup B,\textbf{e}_z)$ can approach this value, which is exactly the case of an infinitely large plate.

According to the proof in the appendix, $2\pi$ is also a supremum when $A$ and $B$ are separated, so a reasonable way to approach this supremum is to mix the mass elements of $A,B$ to approximate a superimposed state. Naively thinking, when mass elements are mixed infinitely fine, the mixed state will always approximate a superimposed state. However, the highly singular integral kernel $K$ makes this intuition false. If we also consider the tidal force in the mixed state as a tensor $T (A,B)= \frac{1}{V}\int_A \int_B K(\textbf{r}-\textbf{r}')\mathrm{d}^3\textbf{r}\mathrm{d}^3\textbf{r}'$, then we get $\Tr T (A,B)= \frac{1}{V}\int_A \int_B 4\pi \delta(\textbf{r} - \textbf{r}')\mathrm{d}^3\textbf{r}\mathrm{d}^3\textbf{r}'=0$.
Since $\Tr T (A,B) = 0$ and $\Tr T (A \cup B) = 2\pi$, the tidal force, as a tensor, differs fundamentally between the mixed and superimposed states, implying that $\Lambda(A,B,\mathbf{n}) \to \Lambda_s(A \cup B, \mathbf{n})$ cannot hold for any $\textbf{n}$ as the mixture becomes infinitely fine.
\begin{figure}[b]
    \centering    
    \hspace*{0.25cm} 
    \vspace*{0cm}
    \includegraphics[width=1\linewidth]{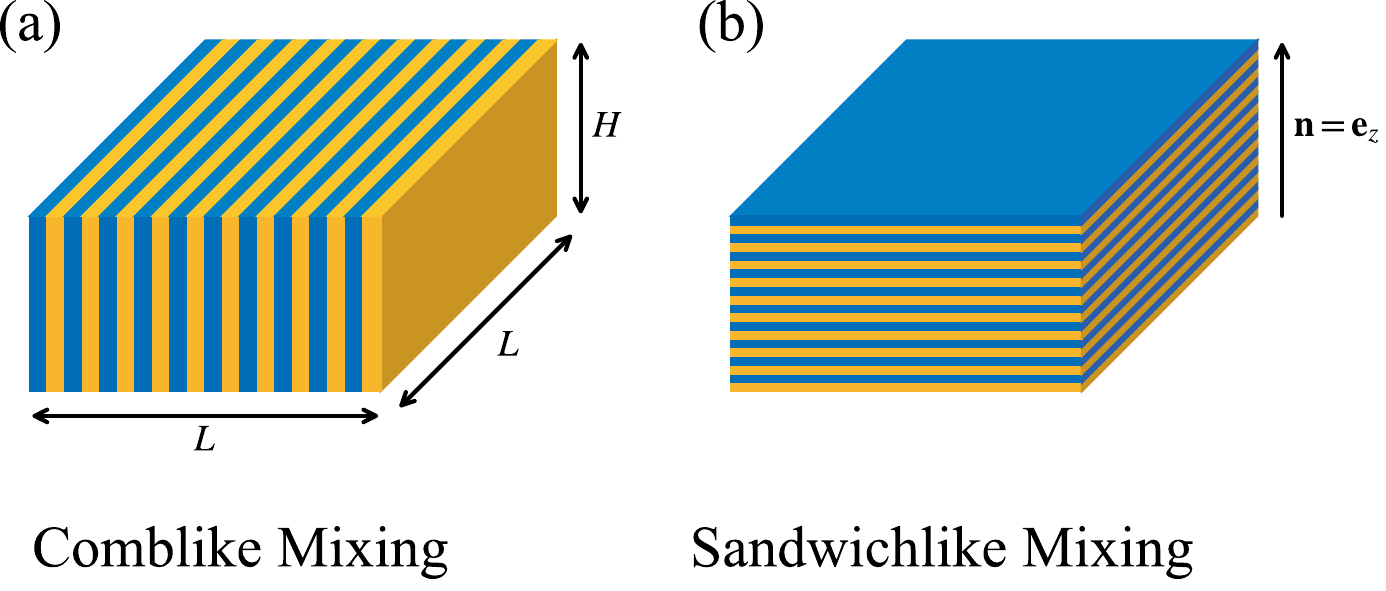}
    \caption{The figure illustrates the dependence of the value of $\Lambda$ on different ways of mixing, even though both converge to the same superimposed state. Fig. \ref{fig:fig2}(a) and Fig. \ref{fig:fig2}(b) show the comblike mixing and sandwichlike mixing of $A \cup B$ as a cuboid of dimensions $L \times L \times H$, with $A$ and $B$ coloured yellow and blue, respectively. The oscillation direction is along the $z$-axis, i.e., $\textbf{n} = \textbf{e}_z$. When the mixture becomes infinitely fine and $H \to 0$, $\Lambda$ approaches $2\pi$ for the former and $0$ for the latter.} 
    \label{fig:fig2}
\end{figure}

Intuitively, one can be convinced that different ways of mixing would result in the same value of $\Lambda$, as long as the configurations converge to the same superimposed state with density $1/2$ when the mixture becomes infinitely fine. However, this conclusion is incorrect due to the singularity of $K$, as the value to which $\Lambda$ converges subtly depends on the way of mixing. Even if two ways of mixing converge to the same superimposed state, they can still yield different values of $\Lambda$. A typical example is shown in Fig. \ref{fig:fig2}, where Figs. \ref{fig:fig2}(a) and \ref{fig:fig2}(b) illustrate two ways of mixing: comb-like mixing and sandwich-like mixing of $A\cup B$ as a $L\times L\times H$ cuboid. As the mixture becomes infinitely fine (i.e., as the number of layers approaches infinity), both cases converge to a superimposed state with a density of $1/2$. However, in the comb-like case, $\Lambda$ approaches $2\pi$ as the number of layers tends to infinity and $H \to 0$, while in the sandwich-like case, $\Lambda$ approaches $0$. Although there are fundamental differences in the tidal force tensors of mixed and superimposed states, as we are only concerned with the tidal force along the $z$-axis, form factor of the mixed state can approximate that of the superimposed state if the  mixture geometry is carefully designed.

One class of correct designs is to make the mixture of $A$ and $B$ translation invariant along the $z$-axis, the height of the plate. Specifically, let $S$ be a large disk on the $xy$-plane, which is the disjoint union of two highly mixed subsets $S_A$ and $S_B$. Now $A=S_A\times [0,H]$, $B=S_B\times [0,H]$, where $H$ is the height of two columns (along $z$-axis). In Eq.~\ref{eq:Lambda}, we can first evaluate the integral of $K_{\textbf{e}_z}(\mathbf{r})$ along $z$-axis from $0$ to $H$ and get
\begin{equation}
	\label{eq:PlaneKernel}
	\begin{aligned}
		\Lambda (A,B,\textbf{e}_z)= \frac{1}{S H}\int_{S_A\times S_B} f(x-x',y-y')\mathrm{d}x\mathrm{d}y\mathrm{d}x'\mathrm{d}y'
	\end{aligned}
\end{equation}
where $f(x,y)=(x^2 +y^2)^{-1/2} - (x^2 +y^2 +H^2)^{-1/2}$. Similarly, the form factor of the superimposed state is  
\begin{equation}
	\label{eq:PlaneKernelSuperImposed}
	\begin{aligned}
		\Lambda_s (A\cup B,\textbf{e}_z) =\frac{1}{4 S H}\int_{S\times S}  f(x-x',y-y')\mathrm{d}x\mathrm{d}y\mathrm{d}x'\mathrm{d}y'
	\end{aligned}
\end{equation}
The previous difficulty does not emerge in this situation because $f(x-x',y-y')$ is less singular than $K$. The singular points of $f$ are $\Delta=\{(x, x', y, y')|x=x',y=y'\}\subset S\times S$. Choosing a small neighbourhood $U$ of $\Delta$, $\int_U f \mathrm{d}x\mathrm{d}y\mathrm{d}x'\mathrm{d}y'$, the integral in $U$ can be arbitrarily small. In $S\times S-U$, $f$ is continuous and bounded, so $\int_{S\times S-U}f \mathrm{d}x\mathrm{d}y\mathrm{d}x'\mathrm{d}y'$ is well defined as a Riemann integral. To be specific, $S\times S-U$ is divided into small cubes and for any small cube $C$, $\int_C f\mathrm{d}x\mathrm{d}y\mathrm{d}x'\mathrm{d}y'$ is approximated by the product of the volume of $C$ and $f(x-x',y-y')$ for some $(x,y,x',y')\in C$. When the mixture becomes infinitely fine, the volume of $(S_A\times S_B)\cap C$ approaches $1/4$ of the volume of $C$, so $4\int_{(S_A\times S_B)\cap C}f \mathrm{d}x\mathrm{d}y\mathrm{d}x'\mathrm{d}y'$ serves as an efficient approximation of $\int_C f\mathrm{d}x\mathrm{d}y\mathrm{d}x'\mathrm{d}y'$ in the Riemann integration. In conclusion, $\Lambda (A,B,\textbf{e}_z)\rightarrow \Lambda_s (S\times [0,H],\textbf{e}_z)$ in an infinitely fine $z$-translation invariant mixture.

\section{Design of oscillators}
\begin{figure}[t]
    \centering    
    \hspace*{0.1cm} 
    \vspace*{0cm}
    \includegraphics[width=1\linewidth]{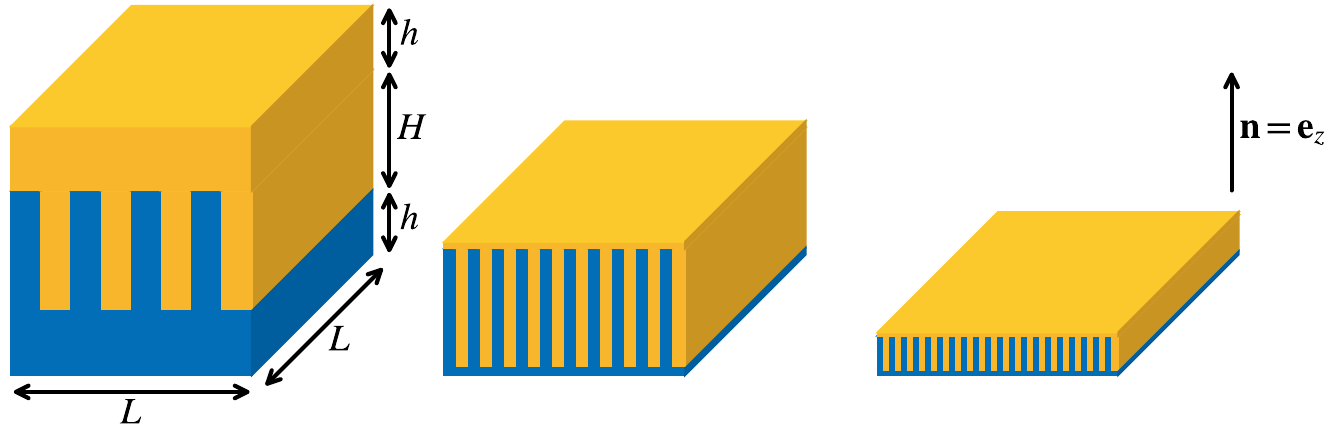}
    \caption{The figure illustrates a sequence of configurations where the form factor tends to $2\pi$. The thickness $h$ of the handles, being independent of other length scales, can decrease arbitrarily quickly such that its contribution can be neglected. As $H \to 0$, the number of teeth $N$ should increase at a faster rate to satisfy $L/N \ll H$.} 
    \label{fig:fig3}
\end{figure}

Now we are able to design an optimal geometry for the two oscillators. They are like two combs, the teeth of which are meshed and form a plate, with each set of teeth fitting alternately with the other. The teeth of combs contribute to the tidal force, while they are glued to two thin handles, in order to make these teeth mechanically connected. Since the thickness $h$ of the handles can always be independently chosen to be much smaller than all other scales, its contribution to $\Lambda$ can be neglected. Based on this, there are only three length scales in the design: the side length $L$ of the plate, the height $H$ of the plate, and the number of teeth $N$, satisfying $L/N \ll H\ll L$. The condition $L/N \ll H$ makes the approximation of the Riemann integral precise, i.e. $\Lambda (A,B,\textbf{e}_z)\rightarrow \Lambda_s (A\cup B,\textbf{e}_z)$, while $H\ll L$ makes $\Lambda_s (A\cup B,\textbf{e}_z)\rightarrow 2\pi$, as seen in Fig. \ref{fig:fig3}. In designing the mixture of objects $A$ and $B$, the translation invariance of teeth along $z$-axis is very important. In contrast, if we use sandwich mixing, with alternating layers stacked on top of each other, $\Lambda(A,B,\textbf{e}_z)$ will tend to zero, as seen in Fig. \ref{fig:fig2}.

Now we do some numerical calculations to verify our arguments. In the following Fig. \ref{fig:fig4}, we vary the number of comb teeth $N$ and plot the value of $\Lambda$ as a function of $H/L$. The patterns in the graph are exactly the same as in our theoretical prediction. First, when $H/L$ is fixed and $N$ is increasing, $\Lambda$ increases and the limit is the value of the superimposed state. Second, when $H/L$ is varying while $N$ is fixed, $\Lambda$ reaches its maximum at a specific point of $H/L$, shown in Table \ref{tab:table1}. This phenomenon is due to the competence of two factors. On the one hand, we need $H\ll L$ to maximize the form factor of the superimposed state; on the other hand, we need $L/N \ll H$ to make the comb approximate the superimposed state. 
\begin{figure}[b]
    \centering    
    \hspace*{-0.8cm} 
    \vspace*{0cm}
    \includegraphics[width=1\linewidth]{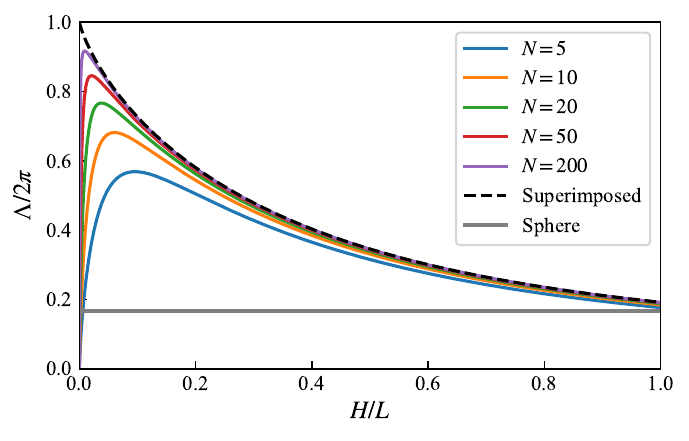}
    \caption{The figure illustrates the form factor $\Lambda$ as a function of $H/L$ for different $N$. As $H \to 0$, $\Lambda$ increases to a maximum value and then decreases to 0. As $N \to \infty$, the maximum value of $\Lambda$ approaches $2\pi$. The black dashed line represents the superimposed case while the grey solid line represents the maximum value of $\Lambda$ for spherical oscillators.
} 
    \label{fig:fig4}
\end{figure}

\begin{table}
\caption{The table shows the height $H$ that maximizes $\Lambda$ for different values of $N$, along with the corresponding values of $\Lambda$.}
\begin{ruledtabular}
\begin{tabular}{cccccccccc} 
	$N$ & $1$ & $2$ & $5$ & $10$ & $20$&$50$&$100$&$200$&$500$ \\ \hline
	$H/L$ & $0.46$  &$0.28$ &$0.16$&$0.10$&$0.07$&$0.04$&$0.03$&$0.02$&$0.01$  \\
	$\Lambda/2\pi$ & $0.19$ & $0.36$ &$0.55$ &$0.66$&$0.75$&$0.83$&$0.87$&$0.91$&$0.94$ \\
\end{tabular}
\end{ruledtabular}
\label{tab:table1}
\end{table}

In experiments, the teeth of two combs should have a gap $a$ to prevent them from touching. Let $a=\epsilon L/2N$, then the teeth have a width of $(1-\epsilon)L/2N$ ($\epsilon\ll1$). Due to the similar argument of Riemann integral, the form factor will be multiplied by a coefficient $1-\epsilon$ as $N \to \infty$. We should choose a suitable $\epsilon$ to make the electrical-magnetic force small enough, compared with gravitational effects.

To maximize the gravity force, oscillators are usually made from metal with large density. The characteristic frequency of mechanical oscillation is usually no more than $\text{kHz}$, where the two oscillators can be treated as perfect conductors. Besides, we can assume they are electrically neutral. In this case, the electrical-magnetic force is mainly from the Casimir effect.

The oscillation is along the height $H$, so we consider the case that the yellow object in Fig. \ref{fig:fig2}(a) has a small translation $z$ at this direction. There are $2N-1$ gaps between these teeth, so the Casimir energy has the form $E_{\text{total}}=(2N-1)E_{\text{single gap}}$.

$E_{\text{single gap}}$ is the Casimir energy for two slabs with gap $a=\epsilon L/2N$, displacement $z$, and side lengths $H$ and $L$. We assume $a=\epsilon L/2N\ll H,L$. 
In this case, we expand the Casimir energy to second order of $z$:

\begin{align}
    E_{\text{single gap}}=\frac{\hbar c\pi^2}{720a^3}HL+\frac{1}{2}kz^2+o(z^2).
\end{align}

The zero-order coefficient of Casimir energy is $\frac{\hbar c\pi^2}{720a^3}HL$ \cite{Casimir:1948dh}. The second-order coefficient is not easy to compute, but for our purpose, it is enough to estimate it using dimensional analysis. As $k$ is irrelevant to $H$ and proportional to $L$, we should have $k=\alpha\frac{\hbar c\pi^2}{720a^4}L$, where $\alpha$ is a dimensionless constant. Therefore, the second-order coefficient of $E_{\text{total}}$ is $(2N-1)\alpha\frac{\hbar c\pi^2}{720}L(\frac{2N}{L\epsilon})^4$.

We require that the Casimir effect be negligible compared to gravitational effects, which implies $(2N)\alpha\frac{\hbar c\pi^2}{720}L(\frac{2N}{L\epsilon})^4\ll GM(2\pi)\rho$. Taking $\rho=22.4\times 10^3 kg/m^3$ for estimation, we get $\frac{HL^5\epsilon^4}{N^5}\gg 6.6\times 10^{-26}\text{m}^6 $. Let $N=1/\epsilon=L/H=10$, which is necessary to distinguish this design from ordinary geometry, it leads to $L\gg 3\text{mm}$. Therefore, Casimir effect can be neglected only when the length scale is larger than centimeter.




\section{Conclusion}
The gravity-induced entanglement (GIE) experiment has recently been considered a strong candidate for testing the quantum nature of gravitational fields. However, one of the significant challenges faced by such experiments is the effects of thermal decoherence in the system. This imposes stringent constraints on the system parameters involved in the experiment. In GIE experiments based on oscillators, this constraint is summarized by the inequality $2\gamma_m k_B T < \hbar G \Lambda \rho$, where $\gamma_m$ is the mechanical dissipation of the oscillator, $T$ is the effective temperature of the system, $\rho$ is the density of the oscillator, and $\Lambda$ is the form factor.
This inequality, arising from the inherent property of the noise model of the GIE, is considered to be universally applicable across diverse experimental systems and cannot be improved with novel quantum control techniques.

In this work, the supremum of the form factor $\Lambda$ for the oscillator is obtained in all geometries and spatial arrangements of the GIE experiment based on the quantum oscillator system. This result provides a fundamental limit for the optimization of system parameters in the future GIE experiments. Compared to the case of the spherical oscillator (for which $\Lambda < \pi/3$), this supremum increases by nearly an order of magnitude, making it particularly important when other system parameters, namely $\gamma_m$, $T$, and $\rho$, are optimized to their limits.

\begin{acknowledgments}
We thank Prof. Haixing Miao for fruitful discussions. Y. Liu acknowledges the support of Beijing Natural Science Foundation (Z240007), National Natural Science Foundation of China (No. 92365210 and No. 12374325), Young Elite Scientists Sponsorship Program by CAST (Grant No. 2023QNRC001), and Beijing Municipal Science and Technology Commission (Grant No. Z221100002722011). Z. Han acknowledges the support of the National Natural Science Foundation of China (Grant No.~T2225008), and the Tsinghua University Dushi Program. Z. T. and H. X contributed equally.
\end{acknowledgments}


\appendix
\section{Proof of Theorem 1}
\label{appendix:A}

This appendix provides the rigorous mathematical proof of Theorem 1 stated in the main text. Recall that the supremum refers to a value such that the variable never exceeds it, but can reach or get arbitrarily close to it, to prove Theorem 1, the key is to show that $\Lambda (A,B,\textbf{n})$ are both $\leq 2\pi$ and can be arbitrarily close to $2\pi$. They can be stated as the following two lemmas. For Lemma 1, we provide the proof in section 1 of Appendix \ref{appendix:A}. And for Lemma 2, the proof is actually already stated in the main text using the properties of Riemann integral. Here we present another more direct proof by giving a construction and evaluating the form factor explicitly in section 2 of Appendix \ref{appendix:A}.

Here, we first present Lemma 1 and Lemma 2, and briefly outline the ideas of their proofs.
\newline

\textit{Lemma 1.-}For any three-dimensional bounded domains $A$ and $B$ of equal volume $V$, where $A$ and $B$ are separated, and any unit vector $\textbf{n}$, the form factor $\Lambda (A,B,\textbf{n})\leq 2\pi$.
\newline

\textit{Lemma 2.-}For any $\epsilon >0$, there exists a pair $(A,B)$ and a direction unit vector $\textbf{n}$ such that
\begin{equation}
    \begin{aligned}
        |\Lambda (A, B, \textbf{n}) - 2\pi|<\epsilon.
    \end{aligned}
\end{equation}

Clearly, by combining Lemma~1 and Lemma~2, we can show that the form factor is not only bounded above by $2\pi$, but can also approach $2\pi$ arbitrarily closely, thereby proving Theorem~1.

To prove Lemma 1, according to the definition of $\Lambda$ (Eq. \ref{eq:Lambda} in the main text), we only need to show that the expression inside the absolute value in Eq. \ref{eq:Lambda} lies between $-2\pi$ and $2\pi$. We first prove, in Lemma 1.1, that this expression can be decomposed into the difference of two parts, each being a semidefinite matrix $\mathcal{M}_i(i=1,2)$ with trace $2\pi$, and contracted on both sides with the unit vector $\textbf{n}$ representing the oscillation direction. With this decomposition, it becomes straightforward to further prove that the expression lies within the range $-2\pi$ to $2\pi$.

To prove Lemma 2, we explicitly constructed a comb-like interleaved configuration (as illustrated in Fig. \ref{fig:fig3}  of the main text) and, through direct calculation, showed that under certain geometric limits, the form factor of this configuration can asymptotically approach $2\pi$. The proof of Lemma 2 mainly consists of four sub-lemmas. In particular, Lemma 2.1 presents the general formula for the form factor of an oscillator composed of two rectangular blocks. Since our constructed configuration can be decomposed into multiple such blocks, we can apply Lemma 2.1 to express its form factor explicitly. Lemmas 2.2 to 2.4 sequentially address three geometric limits: the cap thickness $h$ approaching zero, the number of teeth $N$ tending to infinity, and the height $H$ approaching zero (The definitions of the these geometric parameters can be found in Fig. \ref{fig:fig3} of the main text). Through direct calculation, we demonstrate that by applying these three limits, the form factor asymptotically approaches $2\pi$.

\subsection{Proof of Lemma 1}
\label{subsec:proof_lemma1}

We first introduce the decomposition lemma of the expression inside the absolute value in Eq. \ref{eq:Lambda}:  
\newline

\textit{Lemma 1.1}–For any three-dimensional bounded domains $A$ and $B$ of equal volume $V$, where $A$ and $B$ are separated, and any unit vector $\textbf{n}$, the following integral admits an decomposition:
\begin{equation}
\label{eq:lemma1}
    \begin{aligned}
        &\frac{1}{V}\int_A \int_B \textbf{n}\cdot \nabla_{\textbf{r}}(\textbf{n}\cdot \nabla_{\textbf{r}'}\frac{1}{|\textbf{r} - \textbf{r}'|})\mathrm{d}^3\textbf{r}\mathrm{d}^3\textbf{r}' \\
        &= \textbf{n}^\mathrm{T} \mathcal{M}_1 \textbf{n} - \textbf{n}^\mathrm{T} \mathcal{M}_2 \textbf{n}
    \end{aligned}
\end{equation}
where $\mathcal{M}_i(i=1,2)$ are given by
\begin{equation}
    \begin{aligned}
        \mathcal{M}_i &= \frac{1}{V}\int_{A\cup B} \int_{A\cup B} g_i (\textbf{r}) g_i (\textbf{r}')\nabla_\textbf{r}\otimes \nabla_{\textbf{r}'}\frac{1}{|\textbf{r}-\textbf{r}'|}\mathrm{d}^3\textbf{r}\mathrm{d}^3\textbf{r}'
    \end{aligned}
\end{equation}
while $g_i(i=1,2)$ are given by
\begin{equation}
g_i(\textbf{r}) = \begin{cases} 
1/2 & \text{if } \textbf{r}\in A, \\
(-1)^{i+1}/2 & \text{if } \textbf{r}\in B. 
\end{cases}
\end{equation}
In addition, $\mathcal{M}_i(i=1,2)$ are positive semi-definite matrices with $\Tr \mathcal{M}_1 = \Tr \mathcal{M}_2 = 2\pi$.
\newline

For clarity, let us first explain the notations. Let $f(\mathbf{r}, \mathbf{r}')$ be a scalar function of two coordinates $\textbf{r}=(x,y,z)$ and $\textbf{r}'=(x',y',z')$, then $\nabla_\textbf{r}\otimes \nabla_{\textbf{r}'}f(\mathbf{r}, \mathbf{r}')$ denotes a $3\times 3$ matrix, whose $ij$ component is $\partial_{r_i}\partial_{r'_j}f(\mathbf{r}, \mathbf{r}')$:

\begin{equation}
\label{eq:nabla_tensor}
\begin{aligned}
&\nabla_\textbf{r}\otimes \nabla_{\textbf{r}'}f(\mathbf{r}, \mathbf{r}')\\
=&
\begin{pmatrix}
\partial_x\partial_{x'}f(\mathbf{r}, \mathbf{r}') & \partial_x\partial_{y'}f(\mathbf{r}, \mathbf{r}') & \partial_x\partial_{z'}f(\mathbf{r}, \mathbf{r}') \\
\partial_y\partial_{x'}f(\mathbf{r}, \mathbf{r}') & \partial_y\partial_{y'}f(\mathbf{r}, \mathbf{r}') & \partial_y\partial_{z'}f(\mathbf{r}, \mathbf{r}') \\
\partial_z\partial_{x'}f(\mathbf{r}, \mathbf{r}') & \partial_z\partial_{y'}f(\mathbf{r}, \mathbf{r}') & \partial_z\partial_{z'}f(\mathbf{r}, \mathbf{r}') 
\end{pmatrix}
\end{aligned}
\end{equation}

Therefore, the previously defined $\mathcal{M}_i$ are also $3\times 3$ matrices. By the way, for all multiple integrals appearing in this lemma and subsequent content, it is important to note that integral signs and their corresponding differentials are always paired sequentially from the innermost to the outermost.

This sub-lemma indicates that the expression within the absolute value in Eq. \ref{eq:Lambda}  can be decomposed into the difference of two terms, $\textbf{n}^\mathrm{T} \mathcal{M}_1 \textbf{n}$ and $\textbf{n}^\mathrm{T} \mathcal{M}_2 \textbf{n}$, where $\textbf{n}$ is the oscillation direction vector, and both $\mathcal{M}_1$ and $\mathcal{M}_2$ are positive semi-definite matrices with a trace of $2\pi$. 

To prove Lemma 1.1, we first defines the following matrix $\mathcal{M}$:
\begin{eqnarray}
    \begin{aligned}
        \mathcal{M} &= \frac{1}{V}\int_{A\cup B} \int_{A\cup B} [g_1 (\textbf{r}) - g_2 (\textbf{r})][g_1 (\textbf{r}') + g_2 (\textbf{r}')] \\
        &\nabla_{\textbf{r}}\otimes \nabla_{\textbf{r}'}\frac{1}{|\textbf{r} - \textbf{r}'|}\mathrm{d}^3\textbf{r}\mathrm{d}^3\textbf{r}'.
    \end{aligned}
\end{eqnarray}
Then, $\mathcal{M}$ can be decomposed into four matrices
\begin{eqnarray}
\label{eq:fourmatrices}
    \begin{aligned}
        \mathcal{M} =\mathcal{M}_1 + \mathcal{M}_3 - \mathcal{M}_3^\mathrm{T} - \mathcal{M}_2,
    \end{aligned}
\end{eqnarray}
where $\mathcal{M}_{1,2}$ are already defined in Lemma 1, and
\begin{eqnarray}
    \begin{aligned}
        \mathcal{M}_3 &= \frac{1}{V}\int_{A\cup B} \int_{A\cup B} g_1 (\textbf{r}) g_2 (\textbf{r}')\\
        &\nabla_{\textbf{r}}\otimes \nabla_{\textbf{r}'}\frac{1}{|\textbf{r} - \textbf{r}'|}\mathrm{d}^3\textbf{r}\mathrm{d}^3\textbf{r}'.
    \end{aligned}
\end{eqnarray}
And by definition, the transpose of the $\mathcal{M}_3$ matrix is simply obtained by swapping the two gradients.

With Eq. \ref{eq:fourmatrices}, one obtains
\begin{eqnarray}
\label{eq:M_vs_M1_M2}
    \begin{aligned}
        \textbf{n}^\mathrm{T}\mathcal{M} \textbf{n} &=\textbf{n}^\mathrm{T}\mathcal{M}_1\textbf{n} + \textbf{n}^\mathrm{T}\mathcal{M}_3\textbf{n} - \textbf{n}^\mathrm{T}\mathcal{M}_3^\mathrm{T}\textbf{n} -\textbf{n}^\mathrm{T}\mathcal{M}_2\textbf{n}\\
        &=\textbf{n}^\mathrm{T}\mathcal{M}_1\textbf{n} - \textbf{n}^\mathrm{T}\mathcal{M}_2\textbf{n},
    \end{aligned}
\end{eqnarray}
which is exactly the right-hand side of Eq.\ref{eq:lemma1}.

On the other hand, by definition of $\mathcal{M}$ and by recalling that domains $A$ and $B$ do not overlap, we have:
\begin{eqnarray}
    \begin{aligned}
        \textbf{n}^\mathrm{T}\mathcal{M}\textbf{n}=& \frac{1}{V}\int_{A\cup B} \int_{A\cup B}[g_1 (\textbf{r}) - g_2 (\textbf{r})][g_1 (\textbf{r}') + g_2 (\textbf{r}')]\\
        &\textbf{n}\cdot \nabla_{\textbf{r}}(\textbf{n}\cdot \nabla_{\textbf{r}'}\frac{1}{|\textbf{r} - \textbf{r}'|})\mathrm{d}^3\textbf{r}\mathrm{d}^3\textbf{r}'\\
        =& \frac{1}{V}(\int_{A}\int_{A} + \int_{A}\int_{B} + \int_{B}\int_{A} + \int_{B}\int_{B})\\
        &[g_1 (\textbf{r}) - g_2 (\textbf{r})][g_1 (\textbf{r}') + g_2 (\textbf{r}')]\\
        &\textbf{n}\cdot \nabla_{\textbf{r}}(\textbf{n}\cdot \nabla_{\textbf{r}'}\frac{1}{|\textbf{r} - \textbf{r}'|})\mathrm{d}^3\textbf{r}\mathrm{d}^3\textbf{r}'.\\
    \end{aligned}
\end{eqnarray}

Please note that the first and second integral signs correspond to the integration domains of $\mathbf{r}'$ and $\mathbf{r}$, respectively.
Since the following facts: $g_1 (\textbf{r}) - g_2 (\textbf{r}) = 0$ when $\textbf{r}\in A$, $g_1 (\textbf{r}) - g_2 (\textbf{r}) = 1$ when $\textbf{r}\in B$, $g_1 (\textbf{r}') + g_2 (\textbf{r}') = 1$ when $\textbf{r}'\in A$ and $g_1 (\textbf{r}') + g_2 (\textbf{r}') = 0$ when $\textbf{r}'\in B$, the above integral can be simplified to:
\begin{equation}
    \begin{aligned}
        \textbf{n}^\mathrm{T}\mathcal{M}\textbf{n} &= \frac{1}{V}\int_A \int_B \textbf{n}\cdot \nabla_{\textbf{r}}(\textbf{n}\cdot \nabla_{\textbf{r}'}\frac{1}{|\textbf{r} - \textbf{r}'|})\mathrm{d}^3\textbf{r}\mathrm{d}^3\textbf{r}'.
    \end{aligned}
\end{equation}
This together with Eq. \ref{eq:M_vs_M1_M2} prove the decomposition in Eq. \ref{eq:lemma1}.

To prove the positive semi-definiteness of $\mathcal{M}_i$ for $i=1,2$, one introduces the following functions $\rho_i$ and $\varphi_i$ for $i=1,2$, defined on $\mathbb{R}^3$, given by
\begin{equation}
\label{eq:rhoi}
    \begin{aligned}
        \rho_i (\textbf{r}) = \int_{A\cup B}g_i (\textbf{r}')\textbf{n}\cdot \nabla_{\textbf{r}'}\delta(\textbf{r} -\textbf{r}')\mathrm{d}^3\textbf{r}'
    \end{aligned}
\end{equation}
and
\begin{equation}
\label{eq:phii}
    \begin{aligned}
        \varphi_i (\textbf{r}) = \int_{A\cup B}\frac{\rho_i (\textbf{r}')}{|\textbf{r} - \textbf{r}'|}\mathrm{d}^3\textbf{r}'.
    \end{aligned}
\end{equation}
They can be regarded as two artificially constructed charge distributions and their corresponding potential distributions.

To prove the positive semi-definiteness of $\mathcal{M}_i$, it is sufficient to show that for any unit vector $\mathbf{n}$, $\mathbf{n}^T \mathcal{M}_i \mathbf{n} \geq 0$, which can be seen following:

First, by substituting the definition of $\mathcal{M}_i(i=1,2)$ with only changing the dummy variable of integration to $\textbf{r}''$ and $\textbf{r}'''$, we get:
\begin{equation}
\label{eq:selfenergy}
    \begin{aligned}
        \textbf{n}^\mathrm{T}\mathcal{M}_i\textbf{n}=&\frac{1}{V}\int_{A\cup B} \int_{A\cup B} g_i (\textbf{r}'')g_i (\textbf{r}''')\\
        &\textbf{n}\cdot \nabla_{\textbf{r}''}(\textbf{n}\cdot \nabla_{\textbf{r}'''}\frac{1}{|\textbf{r}'' - \textbf{r}'''|})\mathrm{d}^3\textbf{r}''\mathrm{d}^3\textbf{r}'''.
    \end{aligned}
\end{equation}
By changing the variable in $\nabla_{\textbf{r}'''}$ from $\textbf{r}'''$ to $\textbf{r}''$ (which brings a minus sign), and then using the integral property of the gradient of the delta function: $\int_{\mathbb{R}^3} f(\mathbf{r}) \, \nabla \delta(\mathbf{r} - \mathbf{r}_0) \, d^3\mathbf{r} = - \nabla f(\mathbf{r}_0)$. To avoid ambiguity, we emphasize that in this derivation, the gradient operator immediately to the left of each delta function acts only on the delta function and not on subsequent terms. The expression can be transformed to the following:

\begin{equation}
    \begin{aligned}
        \textbf{n}^\mathrm{T}\mathcal{M}_i\textbf{n}=& \frac{1}{V}\int_{A\cup B} \int_{A\cup B} g_i (\textbf{r}'') g_i(\textbf{r}''')\\
        &\int_{A\cup B} \textbf{n}\cdot \nabla_{\textbf{r}}\delta(\textbf{r} - \textbf{r}'') \textbf{n}\cdot \nabla_{\textbf{r}}\frac{1}{|\textbf{r}-\textbf{r}'''|} {\rm d}^3\textbf{r} {\rm d}^3 \textbf{r}'' {\rm d}^3 \textbf{r}'''.
    \end{aligned}
\end{equation}
By changing the differentiation variable in $\nabla_{\textbf{r}}$ from $\mathbf{r}$ to $\mathbf{r}'''$, an extra negative sign is introduced:

\begin{equation}
    \begin{aligned}
        \textbf{n}^\mathrm{T}\mathcal{M}_i\textbf{n}=&-\frac{1}{V}\int_{A\cup B} \int_{A\cup B} \int_{A\cup B} g_i(\textbf{r}'') g_i(\textbf{r}''')\\
        &\textbf{n}\cdot \nabla_{\textbf{r}}\delta(\textbf{r} - \textbf{r}'') \textbf{n}\cdot \nabla_{\textbf{r}'''}\frac{1}{|\textbf{r}-\textbf{r}'''|} {\rm d}^3\textbf{r} {\rm d}^3 \textbf{r}'' {\rm d}^3 \textbf{r}'''.
    \end{aligned}
\end{equation}
Using the integral property of the gradient of the delta function again, inserting $\nabla_{\textbf{r}'}\delta(\textbf{r}' - \textbf{r}''')$ , the expression is transformed to:
\begin{equation}
    \begin{aligned}
        \textbf{n}^\mathrm{T}\mathcal{M}_i\textbf{n}=& \frac{1}{V}\int_{A\cup B} \int_{A\cup B} \int_{A\cup B} g_i(\textbf{r}'') g_i(\textbf{r}''')\textbf{n}\cdot \nabla_{\textbf{r}}\delta(\textbf{r} - \textbf{r}'')\\
        &\int_{A\cup B} \textbf{n}\cdot \nabla_{\textbf{r}'}\delta(\textbf{r}' - \textbf{r}''')\frac{1}{|\textbf{r}-\textbf{r}'|} {\rm d}^3 \textbf{r}' {\rm d}^3\textbf{r} {\rm d}^3 \textbf{r}'' {\rm d}^3 \textbf{r}'''\\
        =& \frac{1}{V}\int_{A\cup B} \int_{A\cup B} \int_{A\cup B} \int_{A\cup B} g_i(\textbf{r}'') g_i(\textbf{r}''')\\
        &\textbf{n}\cdot \nabla_{\textbf{r}}\delta(\textbf{r} - \textbf{r}'') \textbf{n}\cdot \nabla_{\textbf{r}'}\delta(\textbf{r}' - \textbf{r}''')\frac{1}{|\textbf{r}-\textbf{r}'|}{\rm d}^3\textbf{r} {\rm d}^3 \textbf{r}' {\rm d}^3 \textbf{r}'' {\rm d}^3 \textbf{r}'''.
    \end{aligned}
\end{equation}
Simultaneously changing the variables in both $\nabla$ operators introduces two negative signs, which cancel each other out:
\begin{equation}
    \begin{aligned}
        \textbf{n}^\mathrm{T}&\mathcal{M}_i\textbf{n}= \frac{1}{V}\int_{A\cup B} \int_{A\cup B} \int_{A\cup B} \int_{A\cup B} g_i(\textbf{r}'') g_i(\textbf{r}''')\\
        &\textbf{n}\cdot \nabla_{\textbf{r}''}\delta(\textbf{r} - \textbf{r}'') \textbf{n}\cdot \nabla_{\textbf{r}'''}\delta(\textbf{r}' - \textbf{r}''')\frac{1}{|\textbf{r}-\textbf{r}'|}{\rm d}^3\textbf{r} {\rm d}^3 \textbf{r}' {\rm d}^3 \textbf{r}'' {\rm d}^3 \textbf{r}'''.
    \end{aligned}
\end{equation}
By using the definition of $\rho_i$ and $\varphi_i$ given in Eq. \ref{eq:rhoi} and Eq. \ref{eq:phii}, the expression can be simplified to:

\begin{equation}
    \begin{aligned}
        \textbf{n}^\mathrm{T}\mathcal{M}_i\textbf{n}=& \frac{1}{V}\int_{A\cup B} \int_{A\cup B} \rho_i (\textbf{r}) \rho_i (\textbf{r}')\frac{1}{|\textbf{r}-\textbf{r}'|}{\rm d}^3 \textbf{r}{\rm d}^3\textbf{r}'\\
        =& \frac{1}{V}\int_{A\cup B}\rho_i (\textbf{r}) \varphi_i (\textbf{r})\mathrm{d}^3\textbf{r}\\
        =& \frac{1}{V}\int_{\mathbb{R}^3}\rho_i (\textbf{r}) \varphi_i (\textbf{r})\mathrm{d}^3\textbf{r}.
    \end{aligned}
\end{equation}
By using the Poisson equation $\nabla_{\textbf{r}}^2\varphi_i(\textbf{r})=-4\pi\rho_i(\textbf{r})$ implied by Eq. \ref{eq:phii}, we obtain:

\begin{equation}
    \begin{aligned}
        \textbf{n}^\mathrm{T}\mathcal{M}_i\textbf{n}=& -\frac{1}{4\pi V} \int_{\mathbb{R}^3}\nabla^2_{\textbf{r}}\varphi_i (\textbf{r}) \varphi_i (\textbf{r})\mathrm{d}^3\textbf{r}\\
        =& -\frac{1}{4\pi V} \int_{\mathbb{R}^3}\left(\nabla_{\textbf{r}}[\varphi_i (\textbf{r})\nabla_{\textbf{r}}\varphi_i (\textbf{r})] - [\nabla_{\textbf{r}}\varphi_i (\textbf{r})]^2\right) \mathrm{d}^3\textbf{r}\\
        =&\text{ Surface term} + \frac{1}{4\pi V} \int_{\mathbb{R}^3}[\nabla_{\textbf{r}}\varphi_i (\textbf{r})]^2 \mathrm{d}^3\textbf{r}\geq 0.
    \end{aligned}
\end{equation}
The second equality follows from the gradient formula for the product of functions, while the final equality follows from the Gauss theorem. The final surface term vanishes because all the functions tend to zero at infinity. This finishes the proof of positive semi-definiteness of $\mathcal{M}_i$.

To prove that $\Tr \mathcal{M}_1 = \Tr \mathcal{M}_2 = 2\pi$, we can compute them directly. First, we deal with \( \mathcal{M}_1 \), by successively substituting the definitions of \( \mathcal{M}_1 \) and \( g_1 \), we obtain:
\begin{equation}
    \begin{aligned}
        &\ \ \Tr \mathcal{M}_1\\
        &= \Tr \frac{1}{V}\int_{A\cup B} \int_{A\cup B} g_1(\textbf{r}) g_1(\textbf{r}')\nabla_\textbf{r}\otimes \nabla_{\textbf{r}'}\frac{1}{|\textbf{r} -\textbf{r}'|}\mathrm{d}^3\textbf{r} \mathrm{d}^3\textbf{r}'\\
        &= \Tr \frac{1}{V}\int_{A\cup B} \int_{A\cup B} \frac{1}{4} \nabla_\textbf{r}\otimes \nabla_{\textbf{r}'}\frac{1}{|\textbf{r} -\textbf{r}'|}\mathrm{d}^3\textbf{r} \mathrm{d}^3\textbf{r}'\\
        &= \frac{1}{4V}\int_{A\cup B} \int_{A\cup B} \Tr (\nabla_\textbf{r}\otimes \nabla_{\textbf{r}'}\frac{1}{|\textbf{r} -\textbf{r}'|})\mathrm{d}^3\textbf{r} \mathrm{d}^3\textbf{r}'.
    \end{aligned}
\end{equation}

Recalling the definition for the action of $\nabla_\textbf{r}\otimes \nabla_{\textbf{r}'}$ on a scalar function (Eq. \ref{eq:nabla_tensor}), the trace in the above expression will yield the following result:

\begin{equation}
    \begin{aligned}
        &\ \ \Tr \mathcal{M}_1\\
        &= \frac{1}{4V}\int_{A\cup B} \int_{A\cup B} (\partial_x\partial_{x'}+\partial_y\partial_{y'}+\partial_z\partial_{z'})\frac{1}{|\textbf{r} -\textbf{r}'|}\mathrm{d}^3\textbf{r} \mathrm{d}^3\textbf{r}'.
    \end{aligned}
\end{equation}

Then, based on the symmetry of \( 1/| \mathbf{r} - \mathbf{r'} | \), replacing \( x', y', z' \) in the differential operators with \( x, y, z \) will only introduce an extra minus sign, which leads to:

\begin{equation}
    \begin{aligned}
        &\ \ \Tr \mathcal{M}_1\\
        &= -\frac{1}{4V}\int_{A\cup B} \int_{A\cup B} (\partial_x^2+\partial_y^2+\partial_z^2)\frac{1}{|\textbf{r} -\textbf{r}'|}\mathrm{d}^3\textbf{r} \mathrm{d}^3\textbf{r}'\\
        &= -\frac{1}{4V}\int_{A\cup B} \int_{A\cup B} \nabla_\textbf{r}^2\frac{1}{|\textbf{r} -\textbf{r}'|}\mathrm{d}^3\textbf{r} \mathrm{d}^3\textbf{r}'.
    \end{aligned}
\end{equation}

Using the Poisson equation of a single point charge $\nabla_\textbf{r}^2\frac{1}{|\textbf{r} -\textbf{r}'|}=-4\pi\delta(\textbf{r} - \textbf{r}')$, the expression can be further simplified to:

\begin{equation}
    \begin{aligned}
        &\ \ \Tr \mathcal{M}_1\\
        &= \frac{1}{4V}\int_{A\cup B} \int_{A\cup B} 4\pi \delta(\textbf{r} - \textbf{r}') \mathrm{d}^3\textbf{r} \mathrm{d}^3\textbf{r}'\\
        &= \frac{\pi}{V}\int_{A\cup B} \int_{A\cup B} \delta(\textbf{r} - \textbf{r}') \mathrm{d}^3\textbf{r} \mathrm{d}^3\textbf{r}'\\
        &= \frac{\pi}{V}(\int_{A} \int_{A} + \int_{A} \int_{B} + \int_{B}\int_{A} + \int_{B} \int_{B})\delta(\textbf{r} - \textbf{r}') \mathrm{d}^3\textbf{r} \mathrm{d}^3\textbf{r}'.
    \end{aligned}
\end{equation}

Note that the delta function ensures only when \( \mathbf{r} \) and \( \mathbf{r'} \) coincide contributes to the integral, so only the terms where the integration regions of \( \mathbf{r} \) and \( \mathbf{r'} \) are the same are retained:

\begin{equation}
    \begin{aligned}
        &\ \ \Tr \mathcal{M}_1\\
        &= \frac{\pi}{V}(\int_{A} \int_{A} + \int_{B} \int_{B})\delta(\textbf{r} - \textbf{r}') \mathrm{d}^3\textbf{r} \mathrm{d}^3\textbf{r}'\\
        &= \frac{\pi}{V}(\int_{A}\mathrm{d}^3\textbf{r}'+\int_{B}\mathrm{d}^3\textbf{r}')\\
        &= \frac{\pi}{V}(V+V)\\
        &= 2\pi.
    \end{aligned}
\end{equation}

For $\mathcal{M}_2$, the derivation is similar. We only need to use the definitions of \( \mathcal{M}_2 \) and \( g_2 \), along with the same simplification approach as above.:
\begin{equation}
    \begin{aligned}
        &\ \ \Tr \mathcal{M}_2\\
        &= \Tr \frac{1}{V}\int_{A\cup B} \int_{A\cup B} g_2(\textbf{r})g_2(\textbf{r}')\nabla_\textbf{r}\otimes \nabla_{\textbf{r}'}\frac{1}{|\textbf{r} -\textbf{r}'|}\mathrm{d}^3\textbf{r} \mathrm{d}^3\textbf{r}'\\
        &= \frac{1}{V}\int_{A\cup B} \int_{A\cup B} g_2(\textbf{r})g_2(\textbf{r}')\Tr (\nabla_\textbf{r}\otimes \nabla_{\textbf{r}'}\frac{1}{|\textbf{r} -\textbf{r}'|}) \mathrm{d}^3\textbf{r}'\\
        &= \frac{1}{V}\int_{A\cup B} \int_{A\cup B} g_2(\textbf{r})g_2(\textbf{r}')4\pi \delta(\textbf{r} - \textbf{r}')\mathrm{d}^3\textbf{r} \mathrm{d}^3\textbf{r}'\\
        &= \frac{\pi}{V}\int_{A\cup B}\int_{A\cup B} \delta(\textbf{r} - \textbf{r}')\mathrm{d}^3\textbf{r} \mathrm{d}^3\textbf{r}'\\
        &=\Tr \mathcal{M}_1=2\pi.
    \end{aligned}
\end{equation}
The fourth equality leverages the delta function's property, implying contributing \( \mathbf{r} \) and \( \mathbf{r'} \) must both lie in region $A$ or $B$. Also, from the definition of \( g_2 \), the product of two \( g_2 \) in the same region always equals \( \frac{1}{4} \).
Now we complete the proof of Lemma 1.
\subsection{Proof of Lemma 2}
\label{subsec:proof_lemma2}

To prove Lemma 2, we construct a specific oscillator configuration. Each oscillator consists of a number of mutually parallel rectangular plates (like teeth), which are connected by a cap to form a comb-like object. Two such oscillators are interleaved to form our system. The direction of oscillation is set to be along the $z$-axis. The shape of the oscillator is shown in Fig. \ref{fig:fig3} of the main text. Our proof will first express the form factor of this oscillator configuration, then compute its limiting value under a series of geometric limits, demonstrating that it equals $2\pi$.

The shape of the oscillators we constructed can be rigorously expressed under three-dimensional Cartesian coordinate as the following two domains, $A$ and $B$:

\begin{equation}
    \begin{aligned}
        \overline{A(H,h,N)} &= \{(x,y,z)\in \mathbb{R}^3| \frac{2i-2}{2N}\leq x\leq \frac{2i-1}{2N}\\
        &\text{for some $i=1,...,N$}, 0\leq y\leq 1, 0\leq z\leq H\}\\
        &\cup \{(x,y,z)\in \mathbb{R}^3| 0\leq x,y\leq 1, -h \leq z\leq 0\}\\
       \end{aligned}
\end{equation}
and
\begin{equation}
    \begin{aligned}
        \overline{B(H,h,N)} &= \{(x,y,z)\in \mathbb{R}^3| \frac{2j-1}{2N}\leq x\leq \frac{2j}{2N}\\
        &\text{for some $j=1,...,N$}, 0\leq y\leq 1, 0\leq z\leq H\}\\
        &\cup \{(x,y,z)\in \mathbb{R}^3| 0\leq x,y\leq 1, H \leq z\leq H+h\}.
    \end{aligned}
\end{equation}
Here, $N$ is the number of teeth on each oscillator, $H$ is the height of each tooth, and $h$ is the thickness of the cap. Both the length and width are set to $L = 1$. You may refer directly to Fig. \ref{fig:fig3} in the main text for a visual understanding of these parameters.

Since both $A$ and $B$ are composed of multiple rectangular blocks (both the teeth and the caps), the form factor between $A$ and $B$ can be obtained by summing the contribution corresponding to each pair of rectangular blocks. Therefore, we begin by considering the form factor between an arbitrary pair of rectangular blocks. Without loss of generality, we assume the spatial domains of two rectangular objects are $[x_-, x_+] \times [y_-, y_+] \times [z_-, z_+]$ and $[x'_-, x'_+] \times [y'_-, y'_+] \times [z'_-, z'_+]$, respectively.

In the following Lemma 2.1, we introduce an integral $I$ which equals the integral part of the form factor between such one pair of rectangular blocks (refer to Eq. \ref{eq:Lambda} in the main text, but without dividing by the volume). The integrand in $I$ corresponds to the tidal force between a pair of unit-mass point (with the gravitational constant set to 1). The six-fold integral is taken over the spatial domains of both rectangular blocks.

\textit{Lemma 2.1.} For any $x_i,y_i,z_i,x'_i,y'_i,z'_i\geq 0$ where $i=+,-$, the following sixfold parametrized definite integral can be analytically evaluated
\begin{equation}
    \label{eq:MainIntegral}
    \begin{aligned}
        &I(x_+,x_-,y_+,y_-,z_+,z_-;x'_+,x'_-,y'_+,y'_-,z'_+,z'_-) =\\
        &\int_{x_-}^{x_+}\mathrm{d}x \int_{x'_-}^{x'_+}\mathrm{d}x' \int_{y_-}^{y_+}\mathrm{d}y \int_{y'_-}^{y'_+}\mathrm{d}y' \int_{z_-}^{z_+}\mathrm{d}z \int_{z'_-}^{z'_+}\mathrm{d}z'\\
        &\frac{2(z-z')^2 - (x-x')^2 - (y-y')^2}{[(x-x')^2 + (y-y')^2 + (z-z')^2]^{5/2}}\\
        &=\sum_{i,i',j,j',k,k'\in \{+,-\}}(-1)^{i}(-1)^{i'}(-1)^{j}(-1)^{j'}(-1)^{k}(-1)^{k'}\\
        &F(x_{i}-x_{i'}',y_{j}-y_{j'}',z_{k}-z_{k'}'),
    \end{aligned}
\end{equation}
where $(-1)^\pm = \pm 1$, and
\begin{equation}
    \begin{aligned}
        F(x,y,z) &= \lim_{(x',y',z')\to (x,y,z)}\Bigg[\frac{2z'^2 - x'^2 - y'^2}{6}\sqrt{x'^2 + y'^2 + z'^2}\\
        &+\frac{1}{2}x'(y'^2-z'^2)\tanh^{-1}\frac{x'}{\sqrt{x'^2 + y'^2 + z'^2}}\\
        &+\frac{1}{2}y'(x'^2-z'^2)\tanh^{-1}\frac{y'}{\sqrt{x'^2 + y'^2 + z'^2}}\\
        &-x'y'z'\tan^{-1}\frac{x'y'}{z'\sqrt{x'^2 + y'^2 + z'^2}}\Bigg].
    \end{aligned}
\end{equation}
\newline

The integral in Lemma 2.1 is calculated directly by evaluating the definite integrals layer by layer. One can verify this formula by taking multiple derivatives of the result and checking for consistency. Taking the limit in the expression of $F(x,y,z)$ is  to ensure that the expression is well-defined, even in special cases when some denominators tend to zero.

With the help of Lemma 2.1, for any $H, h>0$ and $N \in \mathbb{N}^+$, one could express the form factor $\Lambda$ of the two comb-like oscillators $A(H,h,N)$ and $B(H,h,N)$ and direction $\textbf{n} =\textbf{e}_z$ in terms of the integral $I$, given by
\begin{equation}
\label{eq:form_lemma2}
    \begin{aligned}
        &\Lambda(A(H,h,N),B(H,h,N),\textbf{e}_z)\\ 
        =& \left|\frac{2}{H+2h}\left[I_1(H,N) + I_2(H,h,N) + I_3(H,h,N) + I_4(H,h)\right]\right|,
    \end{aligned}
\end{equation}
where
\begin{equation}
    \begin{aligned}
        &I_1(H,N)\\ &=\sum_{i,j}I(\frac{2i-1}{2N},\frac{2i-2}{2N},1,0,H,0;\frac{2j}{2N},\frac{2j-1}{2N},1,0,H,0),\\
        \\
        &I_2(H,h,N)\\ &=\sum_{i}^{N}I(\frac{2i-1}{2N},\frac{2i-2}{2N},1,0,H,0;1,0,1,0,H+h,H),\\
        \\
        &I_3(H,h,N)=\sum_{j}^{N}I(\frac{2j}{2N},\frac{2j-1}{2N},1,0,H,0;1,0,1,0,0,-h),\\
        \\
        &I_4(H,h)=I(1,0,1,0,H+h,H;1,0,1,0,0,-h).
    \end{aligned}
\end{equation}
Here, the integral $I_1$ corresponds to the gravitational contribution between the teeth of the two oscillators; integrals $I_2$ and $I_3$ correspond to the gravitational interactions between the teeth of one oscillator and the cap of the other; and $I_4$ accounts for the interaction between the two caps. The prefactor $2 / (H + 2h)$ is the reciprocal of the volume of one oscillator (recall that we have already set $L=1$), matching the $1/V$ factor in Eq. \ref{eq:Lambda} of the main text.

Next, we will take certain limits to verify that the form factor in Eq. \ref{eq:form_lemma2} can approach $2\pi$ arbitrarily closely.


Firstly, fixing any $H$ and $N$, one finds that while $h\to 0$, $I_2, I_3, I_4\to 0$ and only $I_1$ remains. This gives the following Lemma 2.2. This limit corresponds to the case where the cap becomes infinitesimally thin. This conclusion is quite intuitive, since the integrals $I_2$, $I_3$, and $I_4$ all represent the gravitational contributions that involve the caps.
\newline

\textit{Lemma 2.2.}  For any $H>0$ and $N\in\mathbb{N}^+$,
\begin{equation}
    \begin{aligned}
        \lim_{h\to 0^+}\Lambda(A(H,h,N),B(H,h,N),\textbf{e}_z)= \left|\frac{2}{H}I_1(H,N)\right|.
    \end{aligned}
\end{equation}

To rigorously prove Lemma 2.2, it can be seen that $I$ is continuous with respect to its parameters, since $F$ is continuous. Since the parameters of $I$ appear in its limits of integration, its value becomes zero when the two parameters corresponding to the upper and lower limits of the same layer of integration are equal. This gives
\begin{equation}
    \begin{aligned}
        &\lim_{h\to 0^+}I_2(N,h,N)\\
        &= \lim_{h\to 0^+}\sum_{i}^{N}I(\frac{2i-1}{2N},\frac{2i-2}{2N},1,0,H,0;1,0,1,0,H+h,H)\\
        &= \sum_{i}^{N}\lim_{h\to 0^+}I(\frac{2i-1}{2N},\frac{2i-2}{2N},1,0,H,0;1,0,1,0,H+h,H)\\
        &= \sum_{i}^{N}\lim_{h\to 0^+}I(\frac{2i-1}{2N},\frac{2i-2}{2N},1,0,H,0;1,0,1,0,H,H)\\
        &=0.
    \end{aligned}
\end{equation}
For $I_3$ and $I_4$, $\lim_{h\to 0^+}I_{3}(H,h,N)= \lim_{h\to 0^+}I_{4}(H,h) = 0$ can be proved similarly. This proves Lemma 2.2.
\newline

Furthermore, letting $N\to \infty$, which corresponds to making the teeth of the comb sufficiently dense. The following lemma holds, thus the expression in Lemma 2.2 can be further simplified.
\newline

\textit{Lemma 2.3.}  For any $H >0$, define $I_5(H)=I(1,0,1,0,H,0;1,0,1,0,H,0)$, then
\begin{equation}
    \begin{aligned}
        &\lim_{N\to \infty}\sum_{i,j}I_1(H,h,N)=\frac{1}{4}I_5(H).
    \end{aligned}
\end{equation}\newline

To prove Lemma 2.3, we define the following functions
\begin{equation}
    \begin{aligned}
        G(x,x') &= I(x,0,1,0,H,0;x',0,1,0,H,0),\\
        G_{ij}(x,x') &= I(x,\frac{2i-2}{2N},1,0,H,0;x',\frac{2j-1}{2N},1,0,H,0),\\
        g(x,x') &= \frac{\partial^2 G(x,x')}{\partial x\partial x'} =\frac{\partial^2 G_{ij}(x,x')}{\partial x\partial x'}.
    \end{aligned}
\end{equation}
It can be observed that for all $1\leq i,j\leq N$, $i,j\in\mathbb{N}^+$, the function $G_{ij}$ is continuous on $[\frac{2i-2}{2N},\frac{2i-1}{2N}]\times [\frac{2j-1}{2N},\frac{2j}{2N}]$ and is differentiable on $(\frac{2i-2}{2N},\frac{2i-1}{2N})\times (\frac{2j-1}{2N},\frac{2j}{2N})$ by Lemma 2.1. Hence, by Lagrange's mean value theorem, one concludes that for all $1\leq i,j\leq N$, $i,j\in\mathbb{N}^+$, $\exists X_i \in (\frac{2i-2}{2N},\frac{2i-1}{2N})$ and $\exists X'_j \in (\frac{2j-1}{2N},\frac{2j}{2N})$ such that
\begin{equation}
    \begin{aligned}
        &\frac{1}{4N^2}g(X_i,X'_j) = \frac{1}{4N^2}\frac{\partial^2 G_{ij}(x,x')}{\partial x\partial x'}|_{(x,x')=(X_i,X'_j)}\\
        &= I(\frac{2i-1}{2N},\frac{2i-2}{2N},1,0,H,0;\frac{2j}{2N},\frac{2j-1}{2N},1,0,H,0)
    \end{aligned}
\end{equation}
and hence
\begin{equation}
    \begin{aligned}
        &\lim_{N\to \infty}\sum_{i,j}\frac{1}{4N^2}g(X_i,X'_j)\\
        &=\lim_{N\to \infty}\sum_{i,j} I(\frac{2i-1}{2N},\frac{2i-2}{2N},1,0,H,0;\frac{2j}{2N},\frac{2j-1}{2N},1,0,H,0).
    \end{aligned}
\end{equation}
Note that $X_i \in (\frac{2i-2}{2N},\frac{2i-1}{2N})$ and $X'_j \in (\frac{2j-1}{2N},\frac{2j}{2N})$ also implies $X_i \in (\frac{i-1}{N},\frac{i}{N})$ and $X'_j \in (\frac{j-1}{N},\frac{j}{N})$, according to the definition of Riemann integral,
\begin{equation}
    \begin{aligned}
        &\lim_{N\to \infty}\sum_{i,j}\frac{1}{4N^2}g(X_i,X'_j)\\
        &=\frac{1}{4}\lim_{N\to \infty}\sum_{i,j}\frac{1}{N^2}\frac{\partial^2 G(x,x')}{\partial x\partial x'}|_{(x,x')=(X_i,X'_j)}\\
        &=\frac{1}{4}G(1,1) =\frac{1}{4} I(1,0,1,0,H,0;1,0,1,0,H,0)=\frac{1}{4}I_5(H).
    \end{aligned}
\end{equation}
This proves Lemma 2.3.\newline

Finally, the last limit we take is to let the height of the teeth tend to zero. It could be observed that the following lemma holds, and leads us to our final conclusion
\newline

\textit{Lemma 2.4.}
\begin{equation}
    \begin{aligned}
        \lim_{H\to 0^+}\left|\frac{1}{2H}I_5(H)\right| = 2\pi.
    \end{aligned}
\end{equation}\newline

To prove Lemma 2.4, a direct calculation using the expression of $I$ in terms of the function $F$ yields
\begin{equation}
    \begin{aligned}
        &I_5(H)\\
        &=I(1,0,1,0,H,0;1,0,1,0,H,0) \\
        &= 8[F(0,0,0)-F(0,0,H)] - 8[F(0,1,0)-F(0,1,H)] \\
        &- 8[F(1,0,0)-F(1,0,H)] + 8[F(1,1,0)-F(1,1,H)]\\
        &= 8H\tan^{-1}\frac{1}{H\sqrt{2+H^2}} -\frac{8}{3}H^3 \\
        &+ \frac{8}{3}(1-\sqrt{1+H^2}) + \frac{8}{3}(\sqrt{2+H^2}-\sqrt{2})\\
        &+\frac{8}{3}H^2 (2\sqrt{1+H^2} -\sqrt{2+H^2})\\
        &+8H^2 (\tanh^{-1}\frac{1}{\sqrt{2+H^2}} - \tanh^{-1}\frac{1}{\sqrt{1+H^2}})\\
        &+8(\tanh^{-1}\frac{1}{\sqrt{2}} - \tanh^{-1}\frac{1}{\sqrt{2+H^2}}),
    \end{aligned}
\end{equation}
then
\begin{equation}
    \begin{aligned}
        &\lim_{H\to 0^+}\frac{1}{2H}I_5(H)\\
        &=\lim_{H\to 0^+}\frac{1}{2H}I(1,0,1,0,H,0;1,0,1,0,H,0)\\
        &=4\lim_{H\to 0^+}\tan^{-1}\frac{1}{H\sqrt{2+H^2}}\\
        &=4\cdot \frac{\pi}{2}\\
        &=2\pi.
    \end{aligned}
\end{equation}
This proves Lemma 2.4.

Based on Lemma 2.4, for any $\epsilon>0$, there is some $H>0$ such that
\begin{equation}
\label{eq:ineq1}
    \begin{aligned}
        \Bigg|\bigg|\frac{1}{2H}I_5(H)\bigg| - 2\pi\Bigg|<\frac{\epsilon}{3}.
    \end{aligned}
\end{equation}

And based on Lemma 2.3, there is some $N\in \mathbb{N}^+$ such that
\begin{equation}
\label{eq:ineq2}
    \begin{aligned}
        \Bigg| \bigg|\frac{2}{H}I_1(H,N)\bigg|-\left|\frac{1}{2H}I_5(H)\right| \Bigg|\leq & \Bigg|\frac{2}{H}I_1(H,N)-\frac{1}
        {2H}I_5(H)\Bigg|\\
        <&\frac{\epsilon}{3}.
    \end{aligned}
\end{equation}

Finally, based on Lemma 2.2, there is some $h>0$ such that
\begin{equation}
\label{eq:ineq3}
    \begin{aligned}
        &\Bigg|\Lambda(A(H,h,N),B(H,h,N),\textbf{e}_z)-\bigg|\frac{2}{H}I_1(H,N)\bigg|\Bigg|<\frac{\epsilon}{3}.
    \end{aligned}
\end{equation}

By triangle inequality and Eq. \ref{eq:ineq1}-\ref{eq:ineq3}, we arrive at the conclusion that for any $\varepsilon$, there exist some $H$, $h$, and $N$ such that

\begin{equation}
\notag
    \begin{aligned}
        &\Bigg|\Lambda(A(H,h,N),B(H,h,N),\textbf{e}_z) - 2\pi\Bigg|\\
        &= \Bigg|\Lambda(A(H,h,N),B(H,h,N),\textbf{e}_z))-\bigg|\frac{2}{H}I_1(H,N)\bigg|\\
        &+\bigg|\frac{2}{H}I_1(H,N)\bigg|-\bigg|\frac{1}{2H}I_5(H)\bigg|+\bigg|\frac{1}{2H}I_5(H)\bigg|-2\pi\Bigg|\\
        &\leq \Bigg|\Lambda(A(H,h,N),B(H,h,N),\textbf{e}_z))-\bigg|\frac{2}{H}I_1(H,N)\bigg|\Bigg|\\
        &+\Bigg|\bigg|\frac{2}{H}I_1(H,N)\bigg|-\bigg|\frac{1}{2H}I_5(H)\bigg|\Bigg|+\Bigg|\bigg|\frac{1}{2H}I_5(H)\bigg|-2\pi\Bigg|\\
        &<\frac{\epsilon}{3}+\frac{\epsilon}{3}+\frac{\epsilon}{3}\\
        &=\epsilon.
    \end{aligned}
\end{equation}
This completes the proof of Lemma 2.

\nocite{*}

\bibliography{apssamp}

\end{document}